\documentclass[english,aps,preprintnumbers,nofootinbib,superscriptaddress]{revtex4-1}
\usepackage{amssymb,amsmath}

\begin{document}
\preprint{YITP-17-131, IPMU17-0180}
\title{Vector disformal transformation of cosmological perturbations}

\author{Vassilis Papadopoulos}
\email[]{papadopo-AT-clipper.ens.fr}
\affiliation{Ecole Normale Superieure, 45 rue d'Ulm,75005 Paris, France}

\author{Moslem Zarei}
\email[]{m.zarei-AT-cc.iut.ac.ir}

\affiliation{Department of Physics, Isfahan University of Technology, Isfahan
84156-83111, Iran}
\affiliation{ICRANet-Isfahan, Isfahan University of Technology, 84156-83111, Iran}
\affiliation{School of Astronomy, Institute for Research in Fundamental Sciences
(IPM), P. O. Box 19395-5531, Tehran, Iran}

\author{Hassan Firouzjahi}
\email[]{firouz-AT-ipm.ir}
\affiliation{School of Astronomy, Institute for Research in Fundamental Sciences
(IPM), P. O. Box 19395-5531, Tehran, Iran}

\author{Shinji Mukohyama}
\email[]{shinji.mukohyama-AT-yukawa.kyoto-u.ac.jp}
\affiliation{Center for Gravitational Physics, Yukawa Institute for Theoretical Physics, Kyoto University, 606-8502, Kyoto, Japan}
\affiliation{Kavli  Institute  for  the  Physics  and  Mathematics  of  the  Universe  (WPI), The  University  of  Tokyo  Institutes  for  Advanced  Study, The  University  of  Tokyo,  Kashiwa,  Chiba  277-8583,  Japan}
\affiliation{Laboratoire de Math\'ematiques et Physique Th\'eorique (UMR CNRS 7350), Universit\'e Fran\c cois Rabelais, Parc de Grandmont, 37200 Tours, France}

\date{\today}

\begin{abstract}
We study disformal transformations of cosmological perturbations by vector fields in theories invariant under $U(1)$ gauge transformations. Three types of vector disformal transformations are considered: (i) disformal transformations by a single timelike vector; (ii) disformal  transformations by a single spacelike vector; and (iii) disformal transformations by three spacelike vectors. We show that transformations of type (i) do not change either curvature perturbation or gravitational waves; that those of type (ii) do not change curvature perturbation but change gravitational waves; and that those of type (iii) change both curvature perturbation and gravitational waves. Therefore, coupling matter fields to the metric after disformal transformations of type (ii) or (iii) in principle have observable consequences. While the recent multi-messenger observation of binary neutron stars has singled out a proper disformal frame at the present epoch with a high precision, the result of the present paper may thus help distinguishing disformal frames in the early universe.
\end{abstract}

\maketitle

\section{Introduction}

Inflation is arguably the most promising scenario of the early universe, by which
various conceptual issues in the standard big-bang cosmology can be addressed. By
definition it assumes an extended period of accelerated expansion of the universe
followed by a graceful exit, which requires an inflaton field measuring the time
remaining till the end of the accelerated expansion. Thus the system describing
inflation should include not only a metric but also an inflaton field at least, and may
include more fields. The inflaton field may have non-trivial couplings to matter fields, and in
some cases the effects of derivative couplings may be summarized as the existence of
an effective metric to which matter fields are coupled minimally. Out of a metric
$g_{\mu\nu}$ and a scalar field $\phi$ (if the inflaton is a scalar), one can indeed
construct an effective metric of the form
\begin{equation}
 g^{\rm eff}_{\mu\nu} = A(\phi,X)g_{\mu\nu} +
B(\phi,X)\partial_{\mu}\phi\partial_{\nu}\phi\,, \label{eqn:metric-transformation}
\end{equation}
where $X=-g^{\mu\nu}\partial_{\mu}\phi\partial_{\nu}\phi/2$ and $A$ ($>0$) and $B$
are functions of $\phi$ and $X$~\cite{Bekenstein:1992pj}. The transformation from
$g_{\mu\nu}$ to $g^{\rm eff}_{\mu\nu}$ is called conformal if $B=0$. Otherwise, it
is called disformal. While the conformal/disformal transformation is by itself
nothing but a change of variables, coupling matter fields to $g^{\rm eff}_{\mu\nu}$
instead of $g_{\mu\nu}$ may have observable consequences.

One can in principle consider conformal/disformal transformations by a field describing the effects of dark energy. In this case the transformation is again given by
(\ref{eqn:metric-transformation}) but $\phi$ is now considered as a field
responsible for the accelerated expansion of the present universe. Then, the recent
multi-messenger observation of binary neutron stars~ \cite{TheLIGOScientific:2017qsa, GBM:2017lvd}
puts a stringent constraint on the choice of the function $B(\phi,X)$, often called a disformal
factor, if $\partial_{\mu}\phi$ is non-vanishing at the present epoch. Namely, if
$\partial_{\mu}\phi\ne 0$ at the present then the standard model of particle physics
should couple to a frame in which the speed of gravitational waves agrees with the
speed of light within the accuracy of $\mathcal{O}(10^{-15})$~\cite{Monitor:2017mdv}.  In this
sense the observation has singled out a proper disformal frame at the present epoch
with a rather high precision.

On the other hand, a disformal frame in the early universe has not been singled out
by observations so far. One thus might hope that cosmological perturbation might be
a probe of disformal frame in the early universe, especially during inflation, since
it is cosmological perturbation that connects theoretical predictions of
inflationary models with observations. For this reason, many authors have
investigated disformal transformations of the spacetime metric by a scalar field~\cite{Bettoni:2013diz,Minamitsuji:2014waa,Tsujikawa:2014uza,Watanabe:2015uqa,Motohashi:2015pra,Domenech:2015hka,Domenech:2015tca,Fujita:2015ymn}

In the literature there are models of inflation in which not only scalar fields but
also vector fields play important roles, for a review see \cite{review}. This is well-motivated as in theories of high energy physics  and particle physics vector fields (actually gauge fields) play important roles so one naturally expects that  they may play important roles  during inflation as well. Correspondingly,  in those models, vector fields may participate in the disformal transformation between the original metric and the
effective metric. We call such a transformation a vector disformal transformation.

In this work we consider theories invariant under a $U(1)$ gauge transformation so the vector fields are actually the $U(1)$ gauge fields.  We consider three types of vector disformal transformations: (i) disformal transformations by a single timelike vector; (ii) disformal
transformations by a single spacelike vector; and (iii) disformal transformations
by three spacelike vectors. We then investigate the way in which inflationary
cosmological perturbations transform under each of them. We find that
transformations of type (i) do not change either curvature perturbation or
gravitational waves; that those of type (ii) do not change curvature perturbation
but change gravitational waves; and that those of type (iii) change both curvature
perturbation and gravitational waves.

 The rest of the paper is organized as follows. In Sec~\ref{sec:single-timelike} and Sec~\ref{sec:single-spacelike} we investigate disformal transformations of cosmological perturbations by a single timelike vector and a single spacelike vector, respectively. We then consider disformal transformations by three spacelike vectors in Sec~\ref{sec:three-spacelike}. Sec~\ref{sec:summary} is devoted to a summary of the paper and discussions.


\section{A single timelike vector}
 \label{sec:single-timelike}

In our setup we have a complex scalar field $\phi$ which is charged under a $U(1)$ gauge field
$A_\mu$ with the dimensionless charge coupling (electric charge)  $\textbf{e}$.  Our motivation in this work is to extend the previous studies of disformal transformation to models containing  a charged (complex) scalar field which is gauged under a $U(1)$ gauge field. On the physical grounds, one expects the scalar fields in particle physics to be charged under some gauge fields. In addition, symmetry breaking, such as the Higgs mechanism, is a common phenomenon in early universe and high energy physics. Under a $U(1)$ symmetry breaking the gauge field acquires a longitudinal component by eating one degree of freedom from scalar field. In our setup this is achieved by the term  $\textbf{e}^2 |\phi|^2 A_\mu A^\mu$ which generates  an effective  mass term for the gauge field.
 As we shall see, this  effective mass term  plays important roles in  physical properties of the new disformally transformed metric.

With this discussion in mind, we consider  the following disformal transformation
\begin{equation}
\widetilde{g}_{\mu\nu}=g_{\mu\nu}+\lambda \,\textrm{Re}(D_{\mu}\phi\,
\overline{ D_{\nu} \phi })~, \label{eq:initialdisf}
\end{equation}
where the conformal factor is fixed to unity, $\lambda$ is an arbitrary constant
factor and the covariant derivative $D_{\mu}\phi$  is defined as
\begin{equation}
D_{\mu}\phi=\partial_{\mu}\phi+i\, \textbf{e}\phi A_{\mu}~,
\end{equation}
with $\overline{ D_{\mu} \phi }$ denoting its complex conjugation.

We decompose the complex scalar field into the radial and angular parts as follows
\begin{equation}
\phi(x)=\rho(x)e^{i\theta(x)}~.
\end{equation}
Therefore, the disformal metric is simplified to
\begin{equation}
\widetilde{g}_{\mu\nu}=g_{\mu\nu}+\lambda
\left[\,\partial_{\mu}\rho\partial_{\nu}\rho+\rho^2(\partial_{\mu}\theta+\mathbf{e}A_{\mu}))(\partial_{\nu}\theta+\mathbf{e}A_{\nu})\,\right]~. \label{disformal- back0}
\end{equation}
One can check that not only the action and the equations of motion but also the disformal transformation \eqref{disformal- back0} is invariant under the following U(1) gauge transformations
\begin{equation}
\theta\rightarrow\theta+\chi~,\:\:\:\:\:\:\:\:\:\:\:\:\:\:\:A_{\mu}\rightarrow
A_{\mu}-\frac{1}{\mathbf{e}}\,\partial_{\mu}\chi~.
\end{equation}
This freedom allows us to choose the unitary gauge in which  $\theta=0$ and then the disformal transformation is further simplified to
\begin{equation}
\widetilde{g}_{\mu\nu}=g_{\mu\nu}+\lambda
\left[\,\partial_{\mu}\rho\partial_{\nu}\rho+\mathbf{e}^{2}\rho^{2}A_{\mu}A_{\nu}\,\right]~.
\label{disformal- back2}
 \end{equation}

In this section we choose a timelike form of the gauge field at the background level and set
 \begin{equation}
 A_{\mu}(t)=(A_0(t),0,0,0).
\end{equation}
In this background the spacetime retains its SO(3) rotation invariance with the usual isotropic FLRW metric
 \begin{equation}
ds^2=-N(t)^2dt^2 +a(t)^2(dx^2+dy^2+dz^2)~, \label{line0}
\end{equation}
where $N(t)$ is the lapse function and $a(t)$ is the scale factor.

The disformal transformation \eqref{disformal- back2} changes the above line element into the form
\begin{equation}
d\widetilde{s}^{\,2}=-N^2\left(1-\frac{\lambda\,(\dot{\rho}^{\,2}+\textbf{e}^2
\rho^2A_0^2)}{N^2}\right)dt^2+a^2\left(dx^2+dy^2+dz^2\right)~.
\label{disformal-line0}
\end{equation}
We can rewrite the line element \eqref{disformal-line0} in the following
form
\begin{equation}
d\widetilde{s}^{\,2}=-\tilde{N}^2dt^2+a^2\left(dx^2+dy^2+dz^2\right)~,\label{disformal-line1}
\end{equation}
where
\begin{equation}
\tilde{N}=N\left(1-\frac{\lambda\,(\dot{\rho}^{\,2}+\textbf{e}^2
\rho^2A_0^2)}{N^2}\right)^{1/2}~,
\end{equation}
 Therefore, under the timelike vector disformal transformation, the scale factor remains unchanged, while the lapse function changes.

To study the disformal transformation at the perturbation level, we decompose the perturbation of the gauge field as
\begin{eqnarray}
\delta A_{\mu}&=&(\delta A_0, A^{\rm T}_i+\partial_i M)~,
\end{eqnarray}
where $i=x,y,z$; $M$ is a scalar and $A^{\rm T}_i$ is a vector satisfying the transverse condition $\delta^{ij}\partial_{i}A^{\rm T}_j=0$.

The general form of metric perturbation has the following form
\begin{eqnarray}
ds^2=-N^2(1+2\mathcal{A})dt^2+2aN(B_{i}+\partial_{i}B)\,dt
dx^{i}+a^2[(1-2\psi)\delta_{ij}+2\partial_i\partial_jE+\partial_{(i}E_{j)}+h_{ij}]dx^{i}dx^{j}~.\label{pert-line0}
\end{eqnarray}
Here, $\mathcal{A}$, $B$ and $E$ are scalar perturbations, $B_i$, $E_i$ are vector perturbations satisfying the transverse conditions $\delta^{ij}\partial_i B_j=\delta^{ij}\partial_i E_j=0$,  the symmetrization of indices is such that $\partial_{(i}E_{j)}=\partial_iE_j+\partial_jE_i$, and $h_{ij}$ is the transverse traceless tensor perturbation satisfying $\delta^{jk}\partial_kh_{ij}=0=\delta^{ij}h_{ij}$. In appendix~\ref{app:TTpart}, we have presented the prescription to determine the $h_{ij}$ part of the perturbed
metric.

Under the disformal transformation \eqref{disformal- back2}, the perturbed metric transforms as
\begin{equation}
\delta \widetilde{g}_{\mu\nu}=\delta g_{\mu\nu}+\lambda\,\textbf{e}^2
\rho^2(A_{\mu}\delta A_{\nu}+\delta A_{\mu}A_{\nu})
+2\lambda\,\textbf{e}^2A_{\mu}A_{\nu}\rho\delta\rho+\lambda(\partial_{\mu}\rho\partial_{\nu}\delta\rho+\partial_{\mu}\delta\rho\partial_{\nu}\rho)~.\label{tildeg}
\end{equation}
Similarly to the original perturbed metric Eq. (\ref{pert-line0}), we represent the disformally transformed perturbed metric  as
\begin{equation}
d\tilde{s}^2=-\widetilde{N}^2(1+2\widetilde{\mathcal{A}})dt^2+2a\widetilde{N}(\widetilde{B}_{i}+\partial_{i}\widetilde{B})\,dt
dx^{i}+a^2[(1-2\widetilde{\psi})\delta_{ij}+2\partial_i\partial_j\widetilde{E}+\partial_{(i}\widetilde{E}_{j)}+\widetilde{h}_{ij}]dx^{i}dx^{j}~.
\label{dis-pert-metric1}
\end{equation}
 By comparing  the components of the line element $d\widetilde{s}^2$ with
\eqref{tildeg}, we obtain the tilde variables as follows
\begin{eqnarray}
\widetilde{\mathcal{A}}&=&\frac{\mathcal{A}-\lambda(\dot{\rho}\delta\dot{\rho}+\textbf{e}^2\rho^2A_0
\delta
A_0+\textbf{e}^2A_0^2\rho\delta\rho)/N^2}{1-\frac{\lambda(\dot{\rho}^2+\textbf{e}^2\rho^2A_0^2)}{N^2}}~,\:\:\:\:\:\:
\widetilde{\psi}=\psi~,   \:\:\:\:\:\:
\widetilde{E}= E~,    \nonumber \\
\widetilde{B}&=&\frac{B +\lambda(\dot{\rho}\delta\rho+\textbf{e}^2\rho^2
A_0M)/(aN)}{\sqrt{1-\frac{\lambda(\dot{\rho}^2+\textbf{e}^2\rho^2A_0^2)}{N^2}}}~,\:\:\:\:\:\:
\widetilde{B}_{i}=\frac{B_{i}+\lambda\,\textbf{e}^2\rho^2 A_0 A^{\rm T}_i/(aN)
}{\sqrt{1-\frac{\lambda(\dot{\rho}^2+\textbf{e}^2\rho^2A_0^2)}{N^2}}}~,\nonumber \\
\widetilde{E}_i &=& E_i~,\:\:\:\:\:\:
\widetilde{h}_{ij} = h_{ij}~.
\end{eqnarray}

In the present paper we suppose that the system is described by a diffeomorphism invariant theory such as general relativity. This means that physics is invariant under the infinitesimal coordinate transformation
\begin{equation}
x'{}^\mu = x^\mu + \xi^\mu\label{eq:diffeoN} \, ,
\end{equation}
where $\xi^\mu$ is infinitesimal. By taking appropriate combinations of metric perturbations and the perturbations of the  scalar and the gauge field we can define the gauge invariant variables. In appendix~\ref{apsec:gaugetime} we have presented the coordinate transformations of the metric, the scalar field and the gauge field
perturbations in details.

Using the forms of coordinate transformation of these variables,  we find the following gauge invariant scalar and vector variables
\begin{eqnarray}
\delta\rho_{\textrm{\,flat}}&=&\delta\rho+\frac{\dot{\rho}}{N}\frac{\psi}{H}~,\\
\mathcal{A}_{\textrm{flat}}&=&\mathcal{A}+\frac{1}{N}\frac{d}{dt}\left(\frac{\psi}{H}\right)~,\\
\Sigma&=&B-a\frac{\dot{E}}{N}-\frac{1}{a}\frac{\psi}{H}~,\\
\Sigma_i&=&B_i-\frac{a}{N}\dot{E}_i~,
\end{eqnarray}
where $H=\dot a/(aN)$ is the Hubble expansion rate. The tensor variable $h_{ij}$ is gauge-invariant by itself.

In addition,  one finds the following gauge invariant variable associated with the gauge field perturbations
  \begin{eqnarray}
  \delta A_{0\,\textrm{flat}}&=&
  \delta A_{0}+\frac{d}{dt}\left(\frac{A_0}{N}\frac{\psi}{H}\right)~,\\
  M_{\textrm{flat}} &=& M + \frac{A_0}{NH}\psi~,
  \end{eqnarray}
while $A^{\rm T}_i$ is gauge-invariant by itself.

Now, we calculate the gauge transformations of tilde-variables with the new gauge
parameter $\widetilde{\xi}^{\mu}$. The corresponding gauge transformations are also listed in appendix~\ref{apsec:gaugetime}. Using these transformations, one finds the following gauge invariants for the tilde variables
\begin{eqnarray}
\delta\widetilde{\rho}_{\textrm{\,flat}}&=&\delta\rho+ \frac{\dot{\rho}}{\widetilde{N}} \frac{\widetilde{\psi}}{\widetilde{H}}~,\\
\widetilde{\mathcal{A}}_{\textrm{\,flat}}&=&\widetilde{\mathcal{A}} + \frac{1}{\widetilde{N}}\frac{d}{dt}\left(\frac{\widetilde{\psi}}{\widetilde{H}}\right)~,\\
\widetilde{\Sigma}&=&\widetilde{B}-a\frac{\dot{\widetilde{E}}}{\widetilde{N}}-\frac{1}{a}\frac{\widetilde{\psi}}{\widetilde{H}}~ ,\\
\tilde{\Sigma}_i&=&\widetilde{B}_i-\frac{a}{\tilde{N}}\dot{\tilde{E}}_i~,\\
  \delta \tilde{A}_{0\,\textrm{flat}}&=&
  \delta A_{0}+\frac{d}{dt}\left(\frac{A_0}{\tilde{N}}\frac{\tilde{\psi}}{\tilde{H}}\right)~,\\
  \tilde{M}_{\textrm{flat}}& =& M + \frac{A_0}{\tilde{N}\tilde{H}}\tilde{\psi}~,
\end{eqnarray}
where $\widetilde{H}=\dot a/(a\widetilde{N})$. Since $NH=\widetilde{N}\widetilde{H}$ and $\psi=\widetilde{\psi}$, one immediately finds that $\delta\widetilde{\rho}_{\textrm{\,flat}}=\delta\rho_{\textrm{\,flat}}$, $\delta \tilde{A}_{0\,\textrm{flat}}=\delta A_{0\,\textrm{flat}}$ and $\tilde{M}_{\textrm{flat}}=M_{\textrm{flat}}$. Since $A^{\rm T}_i$ does not involve metric perturbation, it is invariant under the disformal transformation, $\tilde{A}^{\rm T}_i=A^{\rm T}_i$. Noting that the  curvature perturbation is proportional to $\delta\widetilde{\rho}_{\textrm{\,flat}}$, we conclude that the curvature perturbation is invariant under this disformal transformation. In addition,  we also have shown above that the tensor perturbation $h_{ij}$ does not change. Therefore, we conclude that the curvature perturbation and tensor perturbations are not affected by the disformal transformations \eqref{disformal- back2}.

On the other hand, we find
\begin{eqnarray}
\widetilde{\mathcal{A}}_{\textrm{\,flat}}&=&\frac{1}{1-\frac{\lambda(\dot{\rho}^2+\textbf{e}^2\rho^2A_0^2)}{N^2}}\left[\mathcal{A}_{\textrm{flat}}
-\frac{\lambda}{N^2}\left(\textbf{e}^2A_0^2\rho\,\delta\rho_{\textrm{\,flat}}+\dot{\rho}\,\delta\dot{\rho}_{\textrm{\,flat}}
+\textbf{e}^2\rho^2A_0\delta A_{0\,\textrm{flat}}\right)\right]~,\\
\widetilde{\Sigma} &=&\frac{\Sigma +\lambda(\dot{\rho}\delta\rho_{\textrm{\,flat}}+\textbf{e}^2\rho^2
A_0M_{\textrm{\,flat}})/(aN)}{\sqrt{1-\frac{\lambda(\dot{\rho}^2+\textbf{e}^2\rho^2A_0^2)}{N^2}}}~,\\
\widetilde{\Sigma}_{i}&=&\frac{\Sigma_{i}+\lambda\,\textbf{e}^2\rho^2 A_0 A^{\rm T}_i/(aN)
}{\sqrt{1-\frac{\lambda(\dot{\rho}^2+\textbf{e}^2\rho^2A_0^2)}{N^2}}}~.
\end{eqnarray}
Therefore, after disformal transformation,  $\mathcal{A}_{\textrm{\,flat}}$, $\Sigma$ and $\Sigma_i$ change depending on $A_0$ and $\dot{\rho}$.

Comparing to the standard perturbation theory with $N=1$, the gauge invariant variable $\Sigma$ is the anisotropic shear. On the other hand, the gauge invariant variable  $\widetilde{\mathcal{A}}_{\textrm{\,flat}}$ is related to the Bardeen potentials $\Phi$ and $\Psi$ \cite{Bassett:2005xm} via
$\widetilde{\mathcal{A}}_{\textrm{\,flat}} = \Phi + \dot \Psi/H$.

 We end this section by concluding that the curvature perturbation is invariant under disformal transformation while the Bardeen potential
$\widetilde{\mathcal{A}}_{\textrm{\,flat}} $ is not. Our result is consistent with that obtained in \cite{Minamitsuji:2014waa,Motohashi:2015pra,Domenech:2015hka} for disformal transformation from timelike derivative of a scalar field.


\section{A single spacelike vector}
 \label{sec:single-spacelike}

 In this section we consider a model of gauge field with a non-vanishing background  spatial component, say along the $x$-direction, where
\begin{equation}
 A_{\mu}(t)=(0,A_x(t),0,0)\,.
 \end{equation}
On the other hand, we suppose that $\rho$ is homogeneous, i.e. $\rho=\rho(t)$, at the background level. This setup is motivated by models of anisotropic inflation in which a background electric field energy density is turned on during inflation. As shown in \cite{Watanabe:2009ct} if one chooses an appropriate coupling between the inflaton field and the gauge field kinetic term the system reaches the attractor regime in which the electric field energy density reaches a small but nearly constant fraction of the total energy density. In this limit, one can obtain a long period of anisotropic inflation which can have interesting observational implications. In particular, these models predict quadrupolar statistical anisotropy in the power spectrum of the curvature perturbation which is constrained by cosmological observations; for a review of related works see   \cite{anisotropic-inflation}.

Since the background is anisotropic the metric is in the form of  Bianchi type I  with the following
line element
\begin{equation}
ds^2=-N(t)^2dt^2 +a(t)^2dx^2+b(t)^2(dy^2+dz^2)~, \label{line}
\end{equation}
where $N(t)$ is the lapse function and $a(t)$ and $b(t)$ are two scale factors. Correspondingly, we have two Hubble expansion rates  $H_a=\dot{a}/(aN)$ and  $H_b=\dot{b}/(bN)$.

The disformal transformation \eqref{disformal- back2} changes the background metric into the the form
\begin{equation}
d\widetilde{s}^{\,2}=-N^2\left(1-\frac{\lambda\,\dot{\rho}^{\,2}}{N^2}\right)dt^2+a^2\left(1+\frac{\lambda\,\textbf{e}^2
\rho^2A_x^2}{a^2}\right)dx^2+b^2\left(dy^2+dz^2\right)~. \label{disformal-line}
\end{equation}
We can rewrite the line element \eqref{disformal-line} in the form
\begin{equation}
d\widetilde{s}^{\,2}=-\tilde{N}^2dt^2+\tilde{a}^2dx^2+b^2\left(dy^2+dz^2\right)~,\label{disformal-line2}
\end{equation}
where
\begin{eqnarray}
\tilde{N}&=&N\left(1-\frac{\lambda\dot{\rho}^{2}}{N^2}\right)^{1/2}~,\label{eqn:tildeN}\\
\tilde{a}&=&a\left(1+\frac{\lambda \,\mathbf{e}^2\rho^2A_x^2}{a^2}\right)^{1/2}~.\label{eqn:tildea}
\end{eqnarray}

 This result shows that the spatial part of the metric in the $y-z$ plane is unchanged under the disformal transformation. However, the lapse function and the component of the metric along the direction of the  background gauge field $A_\mu$ are modified.
 The background gauge field $A_x$
breaks the $SO(3)$ symmetry group down to $SO(2)$. The remaining $SO(2)$
rotational symmetry in the $y-z$ plane allows us to decompose the perturbations into two-dimensional vector and scalar parts. However, in this two-dimensional decomposition, there is no tensor part. Indeed, the three-dimensional tensor perturbations will be restored from the combination of a two-dimensional  scalar and a two-dimensional vector perturbations of the metric when the background becomes isotropic, e.g. at the end of inflation~ \cite{Kanno:2010ab, Emami:2013bk}.

Because of the residual two-dimensional rotational invariance in the $y-z$ plane, we can choose the comoving momentum of modes in Fourier space to have the following form
\begin{equation}
\textbf{k}=(k_x,k_y,0)~.  
\end{equation}

For the scalar and vector parts of the gauge field perturbation, one can
choose respectively \cite{Emami:2013bk}
\begin{eqnarray}
\delta A_{\mu}^{\textrm{scalar}}&=&(\delta A_0,\delta A_x,\partial_y M,0)~,
\\ \delta A_{\mu}^{\textrm{vector}}&=&(0,0,0,D)~ .
\end{eqnarray}

The most general form of the metric perturbation can be written in the following manner
\cite{Himmetoglu:2008ab,Emami:2013bk}
\begin{equation}
\delta g_{\mu\nu}=\left(
                    \begin{array}{ccc}
                      -2N^2\mathcal{A} & aN\beta_{,x} & bN(B_{,p}+B_p)  \\
                       & -2a^2\bar{\psi} & ab(\gamma_{,xp}+\Gamma_{p,x})  \\
                        &  & b^2(-2\psi\delta_{pq}+2E_{,pq}+E_{(p,q)})  \\
                    \end{array}
                  \right)~,\label{gmetric1}
\end{equation}
where $p,q=y, z$. Here, $\mathcal{A}$, $\beta$, $B$, $\overline{}\bar{\psi}$, $\gamma$, $\psi$ and $E$ are scalar perturbations while $B_p$, $E_p$ and $\Gamma_p$ are vectors with respect to rotations in the isotropic $y-z$ plane satisfying the following transverse conditions
\begin{equation}
\delta^{pq}\partial_p B_q=\delta^{pq}\partial_p E_q=\delta^{pq}\partial_p \Gamma_q=0~.
\end{equation}
Therefore, applying these conditions, and noting that $k_z=0$ in our convention, the metric perturbations are represented as
\begin{eqnarray}
\delta g_{\mu\nu}=\left(
                    \begin{array}{cccc}
                      -2N^2\mathcal{A} & aN\beta_{,x} & bNB_{,y} & bN\,B_z \\
                      aN\beta_{,x} & -2a^2\bar{\psi} & ab\gamma_{,xy} &
ab\,\Gamma_{z,x} \\
                       bN\,B_{,y} & ab\gamma_{,xy} & 2b^2(-\psi+E_{,yy}) &
b^2E_{z,y} \\
                      bN\,B_z & ab\,\Gamma_{z,x} & b^2E_{z,y} & -2b^2\,\psi \\
                    \end{array}
                  \right)~.\label{gmetric2}
\end{eqnarray}
The above form of metric perturbations yields the following line element
\begin{eqnarray}
ds^2&=&-N^2(1+2\mathcal{A})d\eta^2+2aN\beta_{,x}\,dt dx+2bNB_{,y}\,dt dy+2bN\,B_z\,dt
dz+a^2(1-2\bar{\psi} )dx^2+2ab\gamma_{,xy}dxdy
 \nonumber \\ && + 2ab\,\Gamma_{z,x}\,dxdz+b^2\left(1-2\psi+2E_{,yy}\right)dy^2+2
b^2E_{z,y}\,dydz +b^2(1-2\psi )\,dz^2~.\label{pert-line}
\end{eqnarray}
On the other hand,  the perturbed parts of the disformal transformation \eqref{disformal- back2} yield \eqref{tildeg}. Therefore, under the disformal transformation, the perturbed metric transforms to
\begin{equation}
\delta \widetilde{g}_{\mu\nu}=\left(
                    \begin{array}{cccc}
                      -2N^2\mathcal{A}+2\lambda\,\dot{\rho}\delta\dot{\rho} &
aN\beta_{,x}+\lambda\,\textbf{e}^2\rho^2A_x\delta
A_0+\lambda\, \dot{\rho}\delta\rho_{,x} &
bNB_{,y}+\lambda\,\dot{\rho}\delta\rho_{,y} & bN\,B_z \\
                      \ast &-2a^2\bar{\psi}+
2\lambda\,\mathbf{e}^2(A_x^2\rho\delta\rho+\rho^2A_x\delta
A_x) &ab\gamma_{,xy}+\lambda\,\mathbf{e}^2\rho^2A_xM_{,y} &
ab\,\Gamma_{z,x}+\lambda\,\mathbf{e}^2\rho^2A_xD \\
                      \ast &  \ast & 2b^2(-\psi+E_{,yy}) &  b^2E_{z,y} \\
                      \ast & \ast & \ast &  -2b^2\,\psi \\
                    \end{array}
                  \right)~.\label{dis-pert-metric1}
\end{equation}
At this step, we rewrite the perturbed line element in the same form as Eq. \eqref{pert-line}
\begin{eqnarray}
d\widetilde{s}^2&=&-\widetilde{N}^2(1+2\widetilde{\mathcal{A}})dt^2+2\widetilde{a}\widetilde{N}\widetilde{\beta}_{,x}\,dt
dx+2b\widetilde{N}\widetilde{B}_{,y}\,dt
dy+2b\widetilde{N}\,\widetilde{B}_z\,dt
dz+\widetilde{a}^2(1-2\widetilde{\bar{\psi}}
)dx^2+2\widetilde{a}b\,\widetilde{\gamma}_{,xy}\,dxdy
 \nonumber \\ && +
2\widetilde{a}b\,\widetilde{\Gamma}_{z,x}\,dxdz+b^2\left(1-2\widetilde{\psi}+2\widetilde{E}_{,yy}\right)dy^2
 +2 b^2\widetilde{E}_{z,y}\,dydz  +b^2(1-2\widetilde{\psi}
)\,dz^2 ~, \label{pert-line-tilde}
\end{eqnarray}
in which $\tilde{N}$ and $\tilde{a}$ are determined at the background level by \eqref{eqn:tildeN} and \eqref{eqn:tildea}.

By comparing the
components of the line element $d\widetilde{s}^2$ with \eqref{dis-pert-metric1}, we
obtain the tilde-variables as follow
\begin{eqnarray}
\widetilde{\psi}&=&\psi~,   \\
\widetilde{E}&=& E~,   \\
\widetilde{E}_z&=& E_z~,   \\
\widetilde{\Gamma}_{z}&=&\frac{\Gamma_{z}+\lambda\,e^2\rho^2A_x\partial^{-1}_{x}D/a}{\sqrt{1+\lambda\,\mathbf{e}^2\rho^2A_x^2/a^2}}~,
\\
\widetilde{\gamma}&=&\frac{\gamma+\lambda\,e^2\rho^2A_x\partial^{-1}_{x}M/a}{\sqrt{1+\lambda\,\mathbf{e}^2\rho^2A_x^2/a^2}}~,\\
\widetilde{\bar{\psi}}&=&\frac{\bar{\psi}- \lambda\,e^2(A_x^2\rho\delta\rho+\rho^2
A_x\delta A_x)/a^2}{1+\lambda\,\mathbf{e}^2\rho^2A_x^2/a^2}~, \\
\widetilde{B}_{z}&=&\frac{B_{z}}{\sqrt{1-\frac{\lambda\,\dot{\rho}^2}{N^2}}}~,\\
\widetilde{B}&=&\frac{1}{\sqrt{1-\frac{\lambda\,\dot{\rho}^2}{N^2}}}\left(B
+\frac{\lambda\,\dot{\rho}\delta\rho}{bN}\right)~,\\
\widetilde{\beta}&=&\frac{\beta+(\lambda\,\textbf{e}^2\rho^2A_x\partial^{-1}_{x}\delta
A_0+\lambda\, \dot{\rho}\delta\rho
)/(aN)}{\sqrt{(1-\frac{\lambda\,\dot{\rho}^2}{N^2})(1+\frac{\lambda\,\textbf{e}^2\rho^2A_x^2}{a^2})}}~,\\
\widetilde{\mathcal{A}}&=&\frac{\mathcal{A}-\lambda\,\dot{\rho}\delta\dot{\rho}/N^2}{1-\frac{\lambda\,\dot{\rho}^2}{N^2}}~.
\end{eqnarray}
As in the previous section, we again take appropriate combinations of the metric perturbations and the perturbations of scalar and gauge fields to define the gauge invariant variables. In appendix~\ref{apsec:gaugeBianchi} we have presented the coordinate transformations of metric, scalar field and gauge field
perturbations.
For the metric defined in \eqref{gmetric2} one  finds the following gauge
invariant scalar variables \cite{Emami:2013bk}
\begin{eqnarray}
\delta\rho_{\textrm{\,flat}}&=&\delta\rho+\frac{\dot{\rho}}{N}\frac{\psi}{H_b}~,\\
\gamma_{\textrm{\,G}}&=&\gamma-\frac{b}{a}E+\frac{a}{b}\,\partial_{x}^{-2}\left(\bar{\psi}-\frac{H_a}{H_b}\psi\right)~,\\
\Gamma_{p\:\textrm{\,G}}&=&\Gamma_p-\frac{b}{a}E_p~,
\end{eqnarray}
where $p=y,z$ and the subscript G indicates that the corresponding variables are gauge invariant.

Correspondingly, one can also calculate the coordinate transformations of the tilde-variables with the new coordinate transformation parameter $\widetilde{\xi}$. The transformations are presented in appendix~\ref{apsec:gaugeBianchi}. Using these transformations, one finds the following gauge invariant variables
\begin{eqnarray}
\delta\widetilde{\rho}_{\textrm{\,flat}}&=&\delta\rho+\frac{\dot{\rho}}{\widetilde{N}}\frac{\widetilde{\psi}}{\widetilde{H}_{b}}~,\\
\widetilde{\gamma}_{\textrm{\,G}}&=&\widetilde{\gamma}-\frac{b}{\widetilde{a}}\widetilde{E}+
\frac{\widetilde{a}}{b}\,\partial_{x}^{-2}(\widetilde{\bar{\psi}}-\frac{\widetilde{H}_a}{\widetilde{H}_b}\widetilde{\psi})~,\\
\widetilde{\Gamma}_{p\:\textrm{\,G}}&=&\widetilde{\Gamma}_p-\frac{b}{\widetilde{a}}\,\widetilde{E}_p~ ,
\end{eqnarray}
where $\widetilde{H}_a=\dot{\widetilde{a}}/(\widetilde{N}\widetilde{a})$ and $\widetilde{H}_b=\dot{b}/(\widetilde{N}b)$. Since $NH_b=\widetilde{N}\widetilde{H}_b$ and $\psi=\widetilde{\psi}$, it is obvious that $\delta\rho_{\textrm{\,flat}}=\delta\widetilde{\rho}_{\textrm{\,flat}}$.

In the following, we are interested in studying the effects of disformal transformation on the gravitational wave (GW) modes. In appendix~\ref{app:TTpart} we provide a closed form expression for the GW modes defined as the transverse-traceless part of the perturbation of the spatial metric. By this expression one can find the GW mode in the 2d decomposition. Indeed, by using \eqref{hIJ}, it is straightforward to extract the transverse-traceless part of the spatial metric perturbation around the Bianchi I background as
\begin{equation}
h_{33}=-\bar{\partial}^{2}\,\bar{\partial}^{-2}_{x}\,h_{22}=-\bar{\partial}^{2}\,\bar{\partial}^{-2}_{y}\,h_{11}=\left(\frac{a}{b}\right)^{1/3}\bar{\partial}^{-2}\,\bar{\partial}^{2}_{y}\,\bar{\partial}^{2}_{x}
\left[\gamma-\frac{b}{a}E +\frac{a}{b}\,\partial^{-2}_{x}(\bar{\psi}-\psi)\right]
~,
\end{equation}
 \begin{equation}
 h_{23}=h_{32}=-\left(\frac{a}{b}\right)^{2/3}\bar{\partial}^{-2}\,\bar{\partial}^{2}_{x}\,\bar{\partial}_{y}\left(\Gamma_z-\frac{b}{a}E_z\right)~,
 \end{equation}
  \begin{equation}
 h_{13}=h_{31}=\left(\frac{a}{b}\right)^{2/3}\bar{\partial}^{-2}\,\bar{\partial}^{2}_{y}\,\bar{\partial}_{x}\left(\Gamma_z-\frac{b}{a}E_z\right)~.
 \end{equation}
 and
 \begin{equation}
h_{12}=h_{21}=\left(\frac{a}{b}\right)^{1/3}\bar{\partial}^{-4}\,\bar{\partial}^{3}_{y}\,\bar{\partial}^{3}_{x}
\left[\gamma-\frac{b}{a}E +\frac{a}{b}\,\partial^{-2}_{x}(\bar{\psi}-\psi)\right]~.
\end{equation}
These results show that $\gamma_{\textrm{\,G}}$ and $\Gamma_{i\:\textrm{\,G}}$ are two GW modes in the isotropic limit, i.e. in the limit $H_a=H_b$.

After realizing the nature of
$\gamma_{\textrm{\,G}}$ and $\Gamma_{i\:\textrm{\,G}}$, we now calculate
$\widetilde{\gamma}_{\textrm{\,G}}$ and $\widetilde{\Gamma}_{i\:\textrm{\,G}}$
in terms of $\gamma_{\textrm{\,G}}$ and $\Gamma_{i\:\textrm{\,G}}$. The
connection between gauge invariant GW modes in both frames help us to realize
how the power spectrum of GW changes under disformal transformations. After
straightforward calculations for $p=z$ we get
\begin{eqnarray}
\widetilde{\gamma}_{{\,\textrm{G}}}
&=&\frac{\gamma_{{\,\textrm{G}}}+\Delta\gamma_{{\,\textrm{G}}}}{\sqrt{1+\frac{\lambda\,\mathbf{e}^2\rho^2A_x^2}{a^2}}}
 ~, \\
\widetilde{\Gamma}_{z G}&=&
\frac{\Gamma_{z G}+\Delta\Gamma_{{z\,\textrm{G}}}}{\sqrt{1+\frac{\lambda\,\mathbf{e}^2\rho^2A_x^2}{a^2}}}
 ~.
\end{eqnarray}
where
\begin{equation}
\Delta\gamma_{G}=-\,
\frac{\lambda\,\mathbf{e}^2\rho^2A_x}{ab}\partial_{x}^{-2}\left(\delta
A_{T}+\frac{A_x}{\rho}\,\delta\rho_{\,{\textrm{flat}}}\right)
\end{equation}
and
\begin{equation}
\Delta\Gamma_{z\,G}=\frac{\lambda\,\mathbf{e}^2\rho^2A_x}{ab}
\partial^{-1}_{x}D ~.
\end{equation}
with
\begin{equation}
\delta A_{T}=\delta A_x-\partial_{x}M+\frac{\dot{A}_x}{NH_b}\psi ~.
\end{equation}
One can check that $\delta A_{T}$, $\Delta\gamma_{G}$
and $\Delta\Gamma_{z G}$ are gauge invariant.

In cosmological observations, one usually considers the correlation functions of the physical variables.
With this motivation in mind, we calculate the power spectra of GW
$\left<\widetilde{\gamma_G}\,\widetilde{\gamma_G}\right>$ and
$\left<\widetilde{\Gamma}_{z G}\,\widetilde{\Gamma}_{z G}\right>$
in new frame.  The results are presented as follows
\begin{eqnarray}
\left<\widetilde{\gamma}_{G}\,\widetilde{\gamma}_{G}\right>&=&\frac{1}{1+\frac{\lambda\,\mathbf{e}^2\rho^2
A_x^2}{a^2}}\Bigg[\,\left<\gamma_{\textrm{\,G}}\,\gamma_{\textrm{\,G}}\right>
+\left(\frac{\lambda\,\mathbf{e}^2\rho^2A_x}{a}\right)^2k_x^{-4}\Big(\left<\delta
A_{T }\,\delta A_{T }\right>
  +\frac{A_x^2}{\rho^{2}}\left<\delta \rho_{\,\textrm{flat}}\,\delta
\rho_{\,\textrm{flat}}\right>
  \nonumber \\ &&
\:\:\:\:\:\:\:\:\:\:\:\:\:\:\:\:\:
+2\frac{A_x}{\rho}\left<\delta A_{T}\,\delta
\rho_{\,\textrm{flat}}\right>\Big)
+2\,\frac{\lambda\,\mathbf{e}^2\rho^2A_x}{ab}k_x^{-2}\Big(\left<\gamma_{\textrm{G}}\,\delta
A_{T }\right>+\frac{A_x}{\rho}\left<\gamma_{\textrm{G}}\,\delta
\rho_{\,\textrm{flat}}\right>
\Big)\,\Bigg]
~,
\end{eqnarray}
and
\begin{equation}
\left<\widetilde{\Gamma}_{z \textrm{\,G}}\,\widetilde{\Gamma}_{z
\textrm{\,G}}\right>=\frac{1}{1+\frac{\lambda\,\mathbf{e}^2\rho^2
A_x^2}{a^2}}\left[\,
\left<\Gamma_{z\,\textrm{\,G}}\,\Gamma_{z\,\textrm{\,G}}\right>+\left(\frac{\lambda\,\mathbf{e}^2\rho^2A_x}{a}\right)^2k_x^{-2}
\left<D\,D\right>\,\right]~.
\end{equation}
The correlation and cross-correlation terms can be obtained by choosing a specific model of  anisotropic
inflation and expanding the action, including the interaction terms, up to second
order, e.g. as in \cite{Kanno:2010ab, Emami:2013bk}.

From these analysis we conclude that, unlike the curvature perturbation, the gravitational waves are affected under the disformal transformation. The key parameter is $\mathbf{e}^2\rho^2A_x^2$ which comes from the ``Higgs mechanism" and which causes the gauge field to acquire a longitudinal mode. Having said this, we further notice that at the end of inflation and reheating the gauge field transfers its energy to radiation, so once the isotropy is restored, we will have $A_x =0$. As a result, the difference under disformal transformation actually disappears and we conclude that the two frames become identical after the restoration of isotropy after inflation.

\section{Three spacelike vectors}
 \label{sec:three-spacelike}

In the previous sections we have set $\theta=0$ by utilizing a gauge freedom. If one
further assumes that
\begin{equation}
 \rho=const.\ne 0\,,
\end{equation}
e.g. by supposing that a $U(1)$ symmetric potential for the complex scalar $\phi$ has
a minimum at $|\phi|\ne 0$ with a sufficiently large mass, then $D_{\mu}\phi=i \mathbf{e} \rho
A_{\mu}$ and the metric $\tilde{g}_{\mu\nu}$ after the disformal transformation
(\ref{eq:initialdisf}) is simply
\begin{equation}
 \tilde{g}_{\mu\nu} = g_{\mu\nu} + \lambda \rho^2 \mathbf{e}^2 A_{\mu}A_{\nu}\,,\label{eq:simpledisf}
\end{equation}
 In this section, for simplicity we start with this form of the vector disformal transformation and then extend it to the system with three vector fields, $A^a_{\mu}$ ($a=1,2,3$). One may then consider this extended system with three vectors as the $\rho^a=const.\ne 0$ limit of the system with three sets of charged complex scalar fields with $U(1)$ gauge symmetry. This brings us to the following extension of the disformal transformation :
\begin{equation}
\tilde{g}_{\mu\nu}=g_{\mu\nu}+\lambda \rho^2 \mathbf{e}^2 A_\mu^aA_\nu^b\delta_{ab}~.\label{eq:disfchiantN}
\end{equation}

Let us now introduce a background solution on which to perform the perturbative analysis. Because we now have three space-like vectors, we can preserve the isotropy of the metric. Hence, we consider a isotropic FLRW background, already defined in (\ref{line0}).

We must also choose a background for the set of three spacelike vectors which we will denote $A_\mu^a$, where $a$ runs from 1 to 3. This is a little less trivial as we must in principle find a solution that fits the metric (\ref{line0}). However, this was done in \cite{Emami:2016ldl} so we will set ourselves within the inflation solution that was constructed in the paper. Thus, at the background level we impose
\begin{equation}
 A_0^a = 0, \quad A_i^a = A(t)\delta^a_i\,,
\end{equation}
which we notice is a set up that preserves isotropy in the physical space. Then, the transformation (\ref{eq:disfchiantN}) transforms the background (\ref{line0}) into the form :
\begin{eqnarray}
   d\tilde{s}^2&=&-N(t)^2dt^2 +a(t)^2\left(1+\lambda \rho^2 \mathbf{e}^2 \frac{A^2}{a^2}\right)\left(dx^2+dy^2+dz^2\right) \nonumber\\
   &=&-N(t)^2dt^2+\tilde{a}(t)^2\left(dx^2+dy^2+dz^2\right)\,,
\end{eqnarray}
where
\begin{equation}
 \tilde{a} = a\sqrt{1+q}\,,\quad q = \lambda \rho^2 \mathbf{e}^2 \frac{A^2}{a^2}\,.
\end{equation}
Hence, at the background level, the disformal transformation only affects the scale factor. As a consequence, the Hubble expansion rate after the disformal transformation is
\begin{equation}
 \tilde{H} = \frac{1}{N}\frac{\dot{\tilde{a}}}{\tilde{a}} = \frac{1}{1+q}\left(H+q\frac{\dot{A}}{NA}\right)\,.
\end{equation}

We can now proceed to introduce linear perturbations.
For the metric, we take the standard perturbations of FLRW as in (\ref{pert-line0}). For the vectors $A_\mu^a$ we take the following perturbations as defined in \cite{Emami:2016ldl}:
\begin{eqnarray}
\delta A^a_0 &=& \delta^{ai}(Y_i+\partial_i Y)\,, \nonumber\\
\delta A^a_i &= &A(t)[ Q\delta^a_i+\partial_i(M_j+\partial_jM)\delta^{aj}+\epsilon_{ijk}\delta^{aj}\delta^{kl}(U_l+\partial_lU)+t_{ij}\delta^{aj}]\,, \label{eqn:deltaAamu}
\end{eqnarray}
where the vector and tensor perturbation variables satisfy the transverse and traceless condition, namely $0=\delta^{ij}\partial_iB_j=\delta^{ij}\partial_iE_j=\delta^{ij}\partial_ih_{jk}=\delta^{ij}\partial_iM_j=\delta^{ij}\partial_iU_j=\delta^{ij}\partial_iY_j=\delta^{ij}\partial_it_{jk}=\delta^{ij}h_{ij} = \delta^{ij}t_{ij}$. We are now ready to investigate the effect of the disformal transformation on the perturbation variables that we defined. We express the new metric $\tilde{g}_{\mu\nu}$ by injecting in the disformal transformation (\ref{eq:disfchiantN}) the expression of the perturbed metric (\ref{pert-line0}) and the perturbed vectors (\ref{eqn:deltaAamu}), while keeping terms up to first order in perturbations. We then obtain the following expression :
\begin{eqnarray}
d\tilde{s}^2 & = & \tilde{g}_{\mu\nu}dx^\mu dx^\nu \nonumber\\
& = & -N^2(1+2 \mathcal{A})dt^2+2aN\left[
\left(B_i+\frac{a}{N}\frac{q}{A}Y_i\right)+\partial_i\left(B+\frac{a}{N}\frac{q}{A}Y\right)\right]dx^idt\nonumber\\
& & +a^2\left[\left(1+q-2\psi+2qQ\right)\delta_{ij}+2\partial_i\partial_j\left(E+qM\right)+\partial_{(i}E_{j)}+q\partial_{(i}M_{j)}+(h_{ij}+2qt_{ij})\right]dx^idx^j \nonumber\\
& = & -N^2(1+2\mathcal{A})dt^2+2\tilde{a} N(\partial_i \tilde{B}+\tilde{B}_i)dx^idt+\tilde{a}^2 \left[(1-2\tilde{\psi})\delta_{ij}+2\partial_i\partial_j\tilde{E}+\partial_{(i}\tilde{E}_{j)}+\tilde{h}_{ij}\right]dx^idx^j\,,
\end{eqnarray}
where, while $\mathcal{A}$ is invariant under the disformal transformation, other components of metric perturbations transform as
\begin{eqnarray}
&&
\tilde{\psi} = \frac{\psi-qQ}{1+q}\,,\qquad
\tilde{E} = \frac{E+qM}{1+q}\,,\qquad
\tilde{E}_i = \frac{E_i+qM_i}{1+q}\,,\nonumber\\
&&
\tilde{B} = \frac{B+\frac{a}{N}\frac{q}{A}Y}{\sqrt{1+q}}\,,\qquad
\tilde{B}_i = \frac{B_i+\frac{a}{N}\frac{q}{A}Y_i}{\sqrt{1+q}}\,,\qquad
\tilde{h}_{ij}=\frac{h_{ij}+2qt_{ij}}{1+q}\,.
\label{ar:perttransfedN}
\end{eqnarray}
Perturbations of the four-vectors $A^a_\mu$, given by (\ref{eqn:deltaAamu}), are not affected by the disformal transformation. As we shall see below, however, the corresponding gauge-invariant variables are not necessarily invariant under the disformal transformation.

Using the transformation laws derived in Appendix~\ref{ap:gauge3space}, we can construct the following set of gauge invariant variables:
\begin{eqnarray}
&&
\Sigma_i=B_i-\frac{a}{N}\dot{E}_i\,,\qquad
Y^{\textrm{flat}}_i=Y_i-A \dot{E}_i\,,\qquad
M^{\textrm{flat}}_i=M_i-E_i\,,\qquad
\mathcal{A}_{\textrm{flat}}=\mathcal{A}+\frac{1}{N}\frac{d}{dt}\left(\frac{\psi}{H}\right)\,,\nonumber\\
&&
Q_{\textrm{flat}}= Q +\psi\frac{\dot{A}}{NAH}\,,\qquad
Y^{\textrm{flat}}=Y-A \dot{E}\,,\qquad
M^{\textrm{flat}}=M-E\,,\qquad
\Sigma=B-a\frac{\dot{E}}{N}-\frac{1}{a}\frac{\psi}{H}\,.
\label{arr:gaugeinvN}
\end{eqnarray}

It is easy to see that $h_{ij}$, $t_{ij}$, $U_i$, $U$ are gauge-invariant by themselves. The number of independent components in these sets is 18, which is what is expected since our initial perturbation variables carried 22 degrees of freedom and we must subtract the 4 that can be gauged out.

We now  inject (\ref{ar:perttransfedN}) in the variables (\ref{arr:gaugeinvN}) to find how the gauge invariant variables transform :
\begin{eqnarray}
&&
\tilde{U}_i=U_i\,, \quad 
\tilde{U}=U\,,\quad
 \tilde{t}_{ij}=t_{ij}\,, \quad
\tilde{h}_{ij}=\frac{h_{ij}+2qt_{ij}}{1+q}\,,\quad
\tilde{Y}^{\textrm{flat}}_i=Y^{\textrm{flat}}_i-A\frac{d}{dt}\left(\frac{q}{1+q}M^{\textrm{flat}}_i\right)\,,\quad
\tilde{M}^{\textrm{flat}}_i=\frac{M^{\textrm{flat}}_i}{1+q}\,, \nonumber\\ 
&&
\tilde{\Sigma}_i=\frac{1}{\sqrt{1+q}}\left[\Sigma_i+\frac{a}{N}\frac{q}{A}Y^{\textrm{flat}}_i-\frac{a}{N}(1+q)\frac{d}{dt}\left({\frac{q}{1+q}}M^{\textrm{flat}}_i\right)\right]\,, \quad
\tilde{\mathcal{A}}_{\textrm{flat}}=\mathcal{A}_{\textrm{flat}}-\frac{1}{N}\frac{d}{dt}{\left(\frac{q}{H+q\frac{\dot{A}}{NA}} Q_{\textrm{flat}}\right)}\,, \nonumber\\
&&
\tilde{Q}_{\textrm{flat}}=\frac{Q_{\textrm{flat}}}{(1+q\frac{\dot{A}}{NAH})}\,,\quad
\tilde{Y}^{\textrm{flat}}=Y^{\textrm{flat}}-A\frac{d}{dt}\left(\frac{q}{1+q}M^{\textrm{flat}}\right)\,, \quad
\tilde{M}^{\textrm{flat}}=\frac{M^{\textrm{flat}}}{1+q}\,,\nonumber\\
&&  \tilde{\Sigma}=\frac{1}{\sqrt{1+q}}\left[\Sigma+\frac{a}{N}\frac{q}{A}Y^{\textrm{flat}}-\frac{a}{N}(1+q)\frac{d}{dt}\left({\frac{q}{1+q}}M^{\textrm{flat}}\right)+\frac{q}{a}\frac{Q_{\textrm{flat}}}{H+q\frac{\dot{A}}{NA}}\right]~.\label{ar:gaugetransfedN}
\end{eqnarray}

While the disformal transformation is by itself nothing but a change of variables, coupling matter fields to $\tilde{g}_{\mu\nu}$ instead of $g_{\mu\nu}$ may have observable consequences. The transformation laws given in (\ref{ar:gaugetransfedN}) then gives an insight of the way the actual observables may change. This can tell us if the two frames are distinguishable at the level of perturbation. In the case of a single timelike vector considered in Sec~\ref{sec:single-timelike} and in the case of a timelike derivative of a scalar field considered in the literature~\cite{Minamitsuji:2014waa,Motohashi:2015pra,Domenech:2015hka}, curvature perturbations are invariant under disformal transformations and thus we cannot differentiate which frame we live in this way. Here, for example, tensor perturbation, which essentially are gravitational waves, are not invariant by the disformal transformation. Indeed, they mix with the tensor perturbations of the 4-vectors. Thus, if we have a theory where two disformally related frames are involved, by analyzing gravitational waves measurement we may obtain some insight concerning the frame where matter couples. Of course, beyond that, (\ref{ar:gaugetransfedN}) provides the freedom of choosing a disformal frame that simplifies the computations.

If we compare these results with what was obtained in Section  \ref{sec:single-timelike} we notice the transformations actually contain a few similarities. Indeed, both conserve the isotropy of the metric, but do so in different ways : on the one side by introducing a time-like vector which do not affect the space sector, on the other side by introducing a set of 3 space-like vectors, which affect the space sector but in an isotropic way.

These different effects are most noticeable at background level : in Section \ref{sec:single-timelike} the scale factor $a$ is unaffected while the lapse function changes. Conversely, in this section, the lapse is unaffected while the scale factor changes. As for the perturbation variables, both cases change $\mathcal{A}_{flat}$ and $\Sigma$. However, the tensor perturbations $h_{ij}$ are invariant in the time-like case, while we saw that they are not when we consider 3 space-like vectors. This is due to the fact that, unlike the case we considered in this section, the single time-like vector does not introduce any tensor perturbation with which gravitational waves may mix.

\section{Summary and discussions}
 \label{sec:summary}

In this work we have studied disformal transformation in models containing a complex scalar field which is charged under a $U(1)$ gauge field. As argued before, in theories of high energy physics the scalar fields are expected to be charged under gauge fields. Because of the Higgs mechanism, the gauge field acquires an effective mass by absorbing one scalar degree of freedom. As we have seen, this phenomenon has non-trivial effects for disformal transformation.

We have studied three classes of models. First, we studied disformal transformation in models which contain only a timelike vector field. Since the background is isotropic, the classifications of metric perturbations in terms of the scalar, vector and tensor perturbations are the same as in standard FLRW backgrounds. We have seen that neither the curvature perturbation nor the tensor perturbations are modified under disformal transformation. This is in par with previous studies of disformal transformation containing only (real) scalar fields. 

In second example, we considered disformal transformation in models containing one spacelike vector field.
This example is motivated from models of anisotropic inflation in which a background electric field is turned on during inflation. These models generally predict statistical anisotropies in curvature perturbation power spectrum. Since the background is anisotropic the spacetime metric is in the form of Bianchi type I. One can decompose the metric perturbations as the scalar and vector perturbations under the rotations of  the isotropic two dimensions. In addition, one scalar and one vector degrees of freedom combine to furnish the  two polarizations of the  final tensor perturbations after inflation. We have found that the curvature perturbation  remains invariant under disformal transformation in this example. However, the tensor perturbations are affected under disformal transformation. The key parameter is the effective mass of the gauge field $\mathbf{e}^2\rho^2A_\mu A^\mu$. However, this modification does not survive after inflation when  the background gauge field transfers its energy to radiation and the universe becomes isotropic. Having said this we conclude that, in principle, when one has a spacelike vector field one expects the tensor modes transform non-trivially under the disformal transformation.

In the third example we studied disformal transformation in models containing three spacelike vectors. We assigned to each gauge field an additional index which can play the role of color for each species. This consideration allows to keep the background isotropic so the spacetime metric is in the form of the FLRW metric. We have seen that again the key parameter is the effective mass of the gauge field  which is identified by the parameter $q$ in this example. We have shown that both curvature perturbations (more specifically the gauge invariant scalar perturbations) and the tensor perturbations are modified under the disformal transformation.

We comment that the second and the third examples studied above are the first cases in literature so far
where the tensor perturbations or  curvature perturbations are affected under  disformal transformation. Mathematically, this is because we have considered disformal transformation via the covariant derivative $D_\mu \phi = \partial_\mu + i \mathbf{e} \phi A_\mu$ which is the physical derivative when the scalar field  is charged under the $U(1)$ gauge field. This in turn explains why the quantity  $\mathbf{e}^2\rho^2A_\mu A^\mu$, arising once the gauge field acquires its longitudinal mode after absorbing one scalar degree of freedom, appears non-trivially in disformal transformation of physical quantities.

\section*{Acknowledgements}
SM thanks the LMPT for hospitality. The work of SM was supported by Japan Society for the Promotion of Science (JSPS) Grants-in-Aid for Scientific Research (KAKENHI) No. 17H02890, No. 17H06359, and by World Premier International Research Center Initiative (WPI), MEXT, Japan. M. Z. and H. F. would like to thank YITP for the warm hospitality while this work was in progress.  V. P. would also like to thank the YITP for welcoming him for an internship where this work was carried out.

\appendix

\section{Extraction of $h_{IJ}$}
\label{app:TTpart}

In this appendix we assume that the spatial part of the background metric is of the form
\begin{equation}
 g^{(0)}_{ij} = \bar{a}(t)^2 \delta_{IJ}e^I_{\ i}e^J_{\ j}\,,
\end{equation}
where $i,j=1,2,3$; $I,J=1,2,3$; and all components of the three spatial $1$-forms $e^{I}_{\ i}$ are constant in space. We then introduce the inverse matrix $e_I^{\ i}$ so that $e^I_{\ i}e_J^{\ i}=\delta^I_J$ and $e^I_{\ i}e_I^{\ j}=\delta^j_i$, and define differential operators $\bar{\partial}_I$, $\bar{\partial}^I$ and $\bar{\Delta}$ as $\bar{\partial}_I\equiv e_I^{\ i}\partial_i$, $\bar{\partial}^I\equiv \delta^{IJ}\bar{\partial}_J$ and $\bar{\Delta}=\delta^{IJ}\bar{\partial}_I\bar{\partial}_J$.

For example, the FLRW background metric \eqref{line0} satisfies the above condition with $\bar{a}=a$ and $e^I_{\ i}=\delta^I_i$. The differential operators $\bar{\partial}_I$ and $\bar{\Delta}$ are the standard partial derivative and the laplacian w.r.t. the isotropic comoving coordinates. The Bianchi I background metric \eqref{line} also satisfies the above condition with $\bar{a}=(ab^2)^{1/3}$, $e^1_{\ i}=(a/b)^{2/3}\delta^1_i$, $e^2_{\ i}=(b/a)^{1/3}\delta^2_i$ and $e^3_{\ i}=(b/a)^{1/3}\delta^3_i$. In the FLRW limit of the Bianchi I metric, i.e. in the limit where $a/b$ becomes constant in time (but does not have to become $1$), the differential operators $\bar{\partial}_I$ and $\bar{\Delta}$ reduce to partial derivative and the laplacian w.r.t. the isotropic comoving coordinates.

We introduce the spatial part of the perturbed metric as
\begin{equation}
 g_{ij} =\left[\bar{a}(t)^2 \delta_{IJ} + \delta g_{IJ}\right] e^I_{\ i}e^J_{\ j}\,,
\end{equation}
where
\begin{equation}
\delta g_{IJ}=2\bar{a}(t)^2\left[-\Psi\delta_{IJ}+F_{,IJ}+\frac{1}{2}F_{(I,J)}+\frac{1}{2}h_{IJ}\right]~.
\end{equation}
Here,  $h_{IJ}$ is transverse and traceless, $\bar{\partial}^{I}h_{IJ}=0=\delta^{IJ}h_{IJ}$, and $F_{I}$ is transverse, $\bar{\partial}^{I}F_{I}=0$. One can extract $h_{IJ}$ from $\delta g_{IJ}$ in the following form
\begin{eqnarray}
\bar{a}^2h_{IJ}&=&\delta g_{IJ} -\frac{1}{2}\delta_{IJ}\left(\delta g - \bar{\Delta}^{-1}\bar{\partial}^K\bar{\partial}^L\delta g_{KL}\right)\nonumber\\
&&
+\frac{1}{2}\bar{\partial}_I\bar{\partial}_{J}\bar{\Delta}^{-1}\left(\delta g + \bar{\Delta}^{-1}\bar{\partial}^K\bar{\partial}^L\delta g_{KL}\right)
- \bar{\Delta}^{-1}\bar{\partial}^K\left(\bar{\partial}_I\delta g_{JK}+\bar{\partial}_J\delta
g_{IK}\right)~,\label{hIJ}
\end{eqnarray}
where $\delta g=\delta^{IJ}\delta g_{IJ}$.

Let us assume that the time-space components of the background metric vanish (as in the FLRW \eqref{line0} and Bianchi I backgrounds \eqref{line}).
\begin{equation}
g^{(0)}_{0i} = g^{(0)}_{i0} = 0\,.
\end{equation}
Under the spatial diffeomorphism,
\begin{equation}
 t\to t\,, \quad x^i \to x^i + \delta^{IJ} e_I^{\ i}\bar{\xi}_J\,,
\end{equation}
the spatial metric perturbation $\delta g_{IJ}$ then transforms as
\begin{equation}
\delta g_{IJ} \to \delta g_{IJ} - \bar{\partial}_I\bar{\xi}_J- \bar{\partial}_J\bar{\xi}_I\,.
\end{equation}
It is easy to show that the transverse-traceless part \eqref{hIJ} is invariant under this transformation, provided that $\bar{\Delta}^{-1}$ is uniquely defined by a proper boundary condition. If we further assume that
\begin{equation}
\partial_t g^{(0)}_{ij} \propto g^{(0)}_{ij}\,, \label{eqn:isotropicexpansion}
\end{equation}
then (\ref{hIJ}) is invariant under the full diffeomorphism,
\begin{equation}
 t\to t + \xi^0\,, \quad  x^i \to x^i + \delta^{IJ} e_I^{\ i}\bar{\xi}_J\,.
\end{equation}
The condition \eqref{eqn:isotropicexpansion} is satisfied by the FLRW background \eqref{line0}. On the other hand, the Bianchi I background \eqref{line} satisfies the condition \eqref{eqn:isotropicexpansion} if and only if $H_a=H_b$ (the constant value of $a/b$ does not have to be unity).


\section{Gauge transformations of perturbations around FLRW background with a timelike vector}
\label{apsec:gaugetime}
Here, we review how metric and gauge field components change under coordinate
transformations. Under a general coordinate transformation $x^{\mu}\rightarrow
x^{\mu}+\xi^{\mu}$, the metric perturbations transform as
\begin{equation}
\delta g_{\mu\nu}\rightarrow \delta g_{\mu\nu}-\mathcal{L}_{\xi}g_{\mu\nu}~,\label{ap:gmunuder}
\end{equation}
where $\mathcal{L}_{\xi}$ denotes the Lie derivative with respect to $\xi^{\mu}$
given by
\begin{equation}
\mathcal{L}_{\xi}g_{\mu\nu}=g_{\mu\nu,\alpha}\xi^{\alpha}+g_{\alpha\nu}\xi_{,\mu}^{\alpha}+g_{\alpha\mu}\xi_{,\nu}^{\alpha}~.\label{ap:lieder}
\end{equation}
For the background defined in \eqref{line0} and with decomposition
$\xi^{\mu}=(\xi^0,\delta^{ij}\lambda_{,j}+\xi_{\perp}^{i})$ with $\partial_i\xi^i_{\perp}=0$ one can derive the gauge
transformations of the perturbed metric \eqref{pert-line0} in the following manner
\begin{eqnarray}
{\cal A}&\rightarrow & {\cal A}-\frac{1}{N}\frac{d}{dt}(N\xi^0)~, \label{eqn:gaugetr_FLRW_calA}\\
B&\rightarrow &B+\frac{N}{a}\xi^0-\frac{a}{N}\dot{\lambda}~,\\
B_i &\rightarrow & B_i-\frac{a}{N}\delta_{ij}\dot{\xi}_{\perp}^{j} ~,\\
\psi &\rightarrow &\psi+\frac{\dot{a}}{a}\,\xi^0~, \\
E &\rightarrow & E-\lambda~,\\
E_i &\rightarrow & E_i-\delta_{ij}\xi_{\bot}^{j}~,\\
h_{ij}&\rightarrow & h_{ij}~. \label{eqn:gaugetr_FLRW_hij}
\end{eqnarray}
As for the gauge field perturbations, we have
\begin{equation}
\delta A_{\mu}\rightarrow \delta A_{\mu}-\mathcal{L}_{\xi}A_{\mu}~,
\end{equation}
where $A_{\mu}$ is the timelike background gauge field and
\begin{equation}
\mathcal{L}_{\xi}A_{\mu}=A_{\mu,\alpha}\xi^{\alpha}+A_{\alpha}\xi_{,\mu}^{\alpha}~.
\end{equation}
In our case the gauge field perturbations transform under coordinate transformations
as
\begin{eqnarray}
\delta A_0 &\rightarrow & \delta A_0 -\frac{d}{dt}(A_0\xi^{0}) ~,\\
\delta A_i &\rightarrow & \delta A_i~,\\
M &\rightarrow & M -A_0\xi^{0}~.
\end{eqnarray}
The gauge transformation of scalar field perturbation is given by
\begin{eqnarray}
\delta \rho &\rightarrow & \delta \rho -\dot{\rho}\xi^{0}~.
\end{eqnarray}
One can also check that the tilde metric perturbations
transform as
\begin{eqnarray}
\widetilde{{\cal A}}&\rightarrow &
\widetilde{{ \cal A}}-\frac{1}{\widetilde{N}}\frac{d}{dt}(\widetilde{N}\widetilde{\xi}^0)~,\\
\widetilde{B}&\rightarrow
&\widetilde{B}+\frac{\widetilde{N}}{\widetilde{a}}\widetilde{\xi}^0-\frac{\widetilde{a}}{\widetilde{N}}\dot{\widetilde{\lambda}}~,\\
\widetilde{B}_i &\rightarrow &
\widetilde{B}_i-\frac{\widetilde{a}}{\widetilde{N}}\delta_{ij}\dot{\widetilde{\xi}}_{\perp}^{j}
~,\\
\widetilde{\psi} &\rightarrow
&\widetilde{\psi}+\frac{\dot{\widetilde{a}}}{\widetilde{a}}\,\widetilde{\xi}^0~, \\
\widetilde{E} &\rightarrow & \widetilde{E}-\widetilde{\lambda}~,\\
\widetilde{E}_i &\rightarrow & \widetilde{E}_i-\delta_{ij}\widetilde{\xi}_{\bot}^{j}~,\\
\widetilde{h}_{ij}&\rightarrow & \widetilde{h}_{ij}~,
\end{eqnarray}
with respect to the gauge parameters
$\widetilde{\xi}^{\mu}=(\widetilde{\xi}^0,\delta^{ij}\widetilde{\lambda}_{,j}+\widetilde{\xi}_{\perp}^{i})$ with $\partial_i\widetilde{\xi}_{\perp}^i=0$.

\section{Gauge transformations of perturbations around Bianchi I background with a single spacelike vector}
\label{apsec:gaugeBianchi}

For the Bianchi I background defined in \eqref{line} and with decomposition
$\xi^{\mu}=(\xi^0,\lambda_{,x},\delta^{pq}\Lambda_{,q}+\xi_{\perp}^{p})$ ($p=y,z$) one can derive the
gauge transformations of the perturbed metric \eqref{gmetric1} in the following
manner \cite{Emami:2013bk}
\begin{eqnarray}
{\cal A}&\rightarrow & {\cal A}-\frac{1}{N}\frac{d}{dt}(N\xi^0)~,\\
\beta &\rightarrow &\beta+\frac{N}{a}\xi^0-\frac{a}{N}\dot{\lambda}~,\\
B&\rightarrow &B+\frac{N}{b}\xi^0-\frac{b}{N}\dot{\Lambda}~,\\
B_p &\rightarrow & B_p-\frac{b}{N}\delta_{pq}\dot{\xi}_{\perp}^{q} ~,\\
\bar{\psi}&\rightarrow &\bar{\psi}+\frac{\dot{a}}{a}\xi^0+\lambda_{,xx}~,\\
\gamma &\rightarrow & \gamma-\frac{b}{a}\Lambda-\frac{a}{b}\lambda~,\\
\psi &\rightarrow &\psi+\frac{\dot{b}}{b}\,\xi^0~, \\
E &\rightarrow & E-\Lambda~,\\
\Gamma_p &\rightarrow &\Gamma_p-\frac{b}{a}\delta_{pq}\xi_{\bot}^{q}~,\\
E_p &\rightarrow & E_p-\delta_{pq}\xi_{\bot}^{q}~.
\end{eqnarray}
On the other hand, the gauge field perturbations transform under coordinate
transformations as
\begin{eqnarray}
\delta A_0 &\rightarrow & \delta A_0 -A_x\dot{\lambda}_{,x} ~,\\
\delta A_x &\rightarrow & \delta A_x -\dot{A}_x\xi^0-A_x\lambda_{,xx}~,\\
M_{,y} &\rightarrow & M_{,y} -A_x\lambda_{,xy}~,\\
D &\rightarrow & D ~,
\end{eqnarray}
and the gauge transformation of scalar field perturbation is given by
\begin{eqnarray}
\delta \rho &\rightarrow & \delta \rho -\dot{\rho}\xi^{0}~.
\end{eqnarray}
One can also check that the tilde metric perturbations
transform as
\begin{eqnarray}
\widetilde{{ \cal A}}&\rightarrow &
\widetilde{ {\cal A}}-\frac{1}{\widetilde{N}}\frac{d}{dt}(\widetilde{{N}}\widetilde{\xi}^0)~,\\
\widetilde{\beta} &\rightarrow
&\widetilde{\beta}+\frac{\widetilde{N}}{\widetilde{a}}\widetilde{\xi}^0-\frac{\widetilde{a}}{\widetilde{N}}\dot{\widetilde{\lambda}}~,\\
\widetilde{B}&\rightarrow
&\widetilde{B}+\frac{\widetilde{N}}{\widetilde{b}}\widetilde{\xi}^0-\frac{\widetilde{b}}{\widetilde{N}}\dot{\Lambda}~,\\
\widetilde{B}_p &\rightarrow &
\widetilde{B}_p-\frac{\widetilde{b}}{\widetilde{N}}\delta_{pq}\dot{\xi}_{\perp}^{q} ~,\\
\widetilde{\bar{\psi}}&\rightarrow
&\widetilde{\bar{\psi}}+\frac{\widetilde{\dot{a}}}{\widetilde{a}}\,\widetilde{\xi}^0+\widetilde{\lambda}_{,xx}~,\\
\widetilde{\gamma} &\rightarrow &
\widetilde{\gamma}-\frac{\widetilde{b}}{\widetilde{a}}\widetilde{\Lambda}-\frac{\widetilde{a}}{\widetilde{b}}\widetilde{\lambda}~,\\
\widetilde{\psi} &\rightarrow
&\widetilde{\psi}+\frac{\widetilde{\dot{b}}}{\widetilde{b}}\,\widetilde{\xi}^0~, \\
\widetilde{E} &\rightarrow & \widetilde{E}-\widetilde{\Lambda}~,\\
\widetilde{\Gamma}_p &\rightarrow
&\widetilde{\Gamma}_p-\frac{\widetilde{b}}{\widetilde{a}}\delta_{pq}\widetilde{\xi}_{\bot}^{q}~,\\
\widetilde{E}_p &\rightarrow & \widetilde{E}_p-\delta_{pq}\widetilde{\xi}_{\bot}^{q}~,
\end{eqnarray}
with respect to the gauge parameters
$\widetilde{\xi}^{\mu}=(\xi^0,\lambda_{,x},\delta^{pq}\Lambda_{,q}+\xi_{\perp}^{p})$.

\section{Gauge transformations of perturbations around FLRW background with 3 spacelike vectors}\label{ap:gauge3space}
In this section we summarize the transformations laws of the perturbations of the metric and the three vector fields, under a gauge transformation. We consider the background and the perturbation introduced in section~\ref{sec:three-spacelike}. We then consider an infinitesimal coordinate transformation $x^\mu \rightarrow x^\mu + \xi^\mu$, where $\xi^{\mu}$ is decomposed as in Appendix~\ref{apsec:gaugetime}.  Then, the metric perturbation transform as in (\ref{eqn:gaugetr_FLRW_calA})-(\ref{eqn:gaugetr_FLRW_hij}). As for the gauge field perturbations, we have
\begin{equation}
\delta A^a_{\mu}\rightarrow \delta A^a_{\mu}-\mathcal{L}_{\xi}A^a_{\mu}~,
\end{equation}
and thus
\begin{eqnarray}
&&
Y \rightarrow Y - A \dot{\lambda}\,,\qquad
Y_i \rightarrow Y_i-A \delta_{ij}\dot{\xi}_\perp^j\,, \qquad
\delta Q \rightarrow\delta Q-\xi^0 \frac{\dot{A}}{A}\,,\qquad
M_i \rightarrow M_i-\delta_{ij}\xi_\perp^j\,,\nonumber\\
&&
M \rightarrow M-\lambda\,,\qquad
U \rightarrow U\,,\qquad
U_i \rightarrow U_i\,,\qquad
t_{ij} \rightarrow t_{ij}\,.
\end{eqnarray}



\begin{thebibliography}{99}




\bibitem{Bekenstein:1992pj}
  J.~D.~Bekenstein,
  Phys.\ Rev.\ D {\bf 48}, 3641 (1993)
  [gr-qc/9211017].

\bibitem{TheLIGOScientific:2017qsa}
  B.~P.~Abbott {\it et al.} 
  Phys.\ Rev.\ Lett.\  {\bf 119}, no. 16, 161101 (2017),
  [arXiv:1710.05832 [gr-qc]].




\bibitem{GBM:2017lvd}
  B.~P.~Abbott {\it et al.} 
  Astrophys.\ J.\  {\bf 848}, no. 2, L12 (2017)
  [arXiv:1710.05833 [astro-ph.HE]].


\bibitem{Monitor:2017mdv}
  B.~P.~Abbott {\it et al.} 
  Astrophys.\ J.\  {\bf 848}, no. 2, L13 (2017)
  [arXiv:1710.05834 [astro-ph.HE]].


\bibitem{Bettoni:2013diz}
  D.~Bettoni and S.~Liberati,
  Phys.\ Rev.\ D {\bf 88}, 084020 (2013)
  doi:10.1103/PhysRevD.88.084020
  [arXiv:1306.6724 [gr-qc]].

\bibitem{Minamitsuji:2014waa}
  M.~Minamitsuji,
  Phys.\ Lett.\ B {\bf 737}, 139 (2014)
  [arXiv:1409.1566 [astro-ph.CO]].

\bibitem{Tsujikawa:2014uza}
  S.~Tsujikawa,
  JCAP {\bf 1504}, no. 04, 043 (2015)
  doi:10.1088/1475-7516/2015/04/043
  [arXiv:1412.6210 [hep-th]].

\bibitem{Watanabe:2015uqa}
  Y.~Watanabe, A.~Naruko and M.~Sasaki,
  EPL {\bf 111}, no. 3, 39002 (2015)
  doi:10.1209/0295-5075/111/39002
  [arXiv:1504.00672 [gr-qc]].

  \bibitem{Motohashi:2015pra}
  H.~Motohashi and J.~White,
  JCAP {\bf 1602}, no. 02, 065 (2016)
  [arXiv:1504.00846 [gr-qc]].

\bibitem{Domenech:2015hka}
  G.~Domènech, A.~Naruko and M.~Sasaki,
  JCAP {\bf 1510}, no. 10, 067 (2015)
  doi:10.1088/1475-7516/2015/10/067
  [arXiv:1505.00174 [gr-qc]].

\bibitem{Domenech:2015tca}
  G.~Domènech, S.~Mukohyama, R.~Namba, A.~Naruko, R.~Saitou and Y.~Watanabe,
  Phys.\ Rev.\ D {\bf 92}, no. 8, 084027 (2015)
  doi:10.1103/PhysRevD.92.084027
  [arXiv:1507.05390 [hep-th]].

\bibitem{Fujita:2015ymn}
  T.~Fujita, X.~Gao and J.~Yokoyama,
  JCAP {\bf 1602}, no. 02, 014 (2016)
  doi:10.1088/1475-7516/2016/02/014
  [arXiv:1511.04324 [gr-qc]].

\bibitem{review}
J.~Soda,
Class.\ Quant.\ Grav.\  {\bf 29}, 083001 (2012)
[arXiv:1201.6434 [hep-th]].\\
A.~Maleknejad, M.~M.~Sheikh-Jabbari and J.~Soda,
Phys.\ Rept.\  {\bf 528}, 161 (2013)
[arXiv:1212.2921 [hep-th]],\\
 R.~Emami,
  arXiv:1511.01683 [astro-ph.CO].

\bibitem{Bassett:2005xm}
  B.~A.~Bassett, S.~Tsujikawa and D.~Wands,
  Rev.\ Mod.\ Phys.\  {\bf 78}, 537 (2006),
  [astro-ph/0507632].






\bibitem{Watanabe:2009ct}
  M.~a.~Watanabe, S.~Kanno and J.~Soda,
  Phys.\ Rev.\ Lett.\  {\bf 102}, 191302 (2009)
  [arXiv:0902.2833 [hep-th]].




 \bibitem{anisotropic-inflation}
J.~Ohashi, J.~Soda and S.~Tsujikawa,
JCAP {\bf 1312}, 009 (2013).

J.~Ohashi, J.~Soda and S.~Tsujikawa,
Phys.\ Rev.\ D {\bf 88}, 103517 (2013).

J.~Ohashi, J.~Soda and S.~Tsujikawa,
Phys.\ Rev.\ D {\bf 87}, 083520 (2013).

S.~Kanno, J.~Soda, M.~-a.~Watanabe,
JCAP {\bf 1012}, 024 (2010).

K.~Murata, J.~Soda,
JCAP {\bf 1106}, 037 (2011).

S.~Yokoyama and J.~Soda,
JCAP {\bf 0808}, 005 (2008).

K.~Yamamoto, M.~-a.~Watanabe and J.~Soda,
Class.\ Quant.\ Grav.\  {\bf 29}, 145008 (2012).

  A.~Ito and J.~Soda,
  Phys.\ Rev.\ D {\bf 92}, no. 12, 123533 (2015).

  A.~Ito and J.~Soda,
  JCAP {\bf 1604}, no. 04, 035 (2016).


R.~Emami, H.~Firouzjahi, S.~M.~Sadegh Movahed, M.~Zarei,
JCAP {\bf 1102 } (2011)  005.

R.~Emami and H.~Firouzjahi,
JCAP {\bf 1201}, 022 (2012).


  S.~Baghram, M.~H.~Namjoo and H.~Firouzjahi,
  JCAP {\bf 1308}, 048 (2013).



  R.~Emami and H.~Firouzjahi,
  JCAP {\bf 1510}, no. 10, 043 (2015).


A.~A.~Abolhasani, R.~Emami, J.~T.~Firouzjaee and H.~Firouzjahi,
JCAP {\bf 1308}, 016 (2013).


  A.~A.~Abolhasani, R.~Emami and H.~Firouzjahi,
  JCAP {\bf 1405}, 016 (2014).

  X.~Chen, R.~Emami, H.~Firouzjahi and Y.~Wang,
  JCAP {\bf 1408}, 027 (2014).

  N.~Bartolo, S.~Matarrese, M.~Peloso and A.~Ricciardone,
  Phys.\ Rev.\ D {\bf 87}, 023504 (2013).

  T.~R.~Dulaney and M.~I.~Gresham,
  Phys.\ Rev.\ D {\bf 81}, 103532 (2010),


M.~Shiraishi, E.~Komatsu, M.~Peloso and N.~Barnaby,
JCAP {\bf 1305}, 002 (2013).

  M.~Shiraishi, E.~Komatsu and M.~Peloso,
  JCAP {\bf 1404}, 027 (2014).

  N.~Barnaby, R.~Namba and M.~Peloso,
  Phys.\ Rev.\ D {\bf 85}, 123523 (2012).



  K.~Dimopoulos, M.~Karciauskas, D.~H.~Lyth and Y.~Rodriguez,
  JCAP {\bf 0905}, 013 (2009).


  A.~E.~Gumrukcuoglu, B.~Himmetoglu, M.~Peloso,
  Phys.\ Rev.\  {\bf D81}, 063528 (2010).

K.~Yamamoto,
Phys.\ Rev.\ D {\bf 85}, 123504 (2012).

  H.~Funakoshi and K.~Yamamoto,
  Class.\ Quant.\ Grav.\  {\bf 30}, 135002 (2013).


T.~Fujita and S.~Yokoyama,
JCAP {\bf 1309}, 009 (2013).

  T.~Fujita and I.~Obata,
  arXiv:1711.11539 [astro-ph.CO].



S.~R.~Ramazanov and G.~Rubtsov,
Phys.\ Rev.\ D {\bf 89}, 043517 (2014).


  S.~Nurmi and M.~S.~Sloth,
  JCAP {\bf 1407}, 012 (2014).


  R.~K.~Jain and M.~S.~Sloth,
  JCAP {\bf 1302}, 003 (2013).


  F.~R.~Urban,
  Phys.\ Rev.\ D {\bf 88}, 063525 (2013).


M.~Thorsrud, D.~F.~Mota and S.~Hervik,
JHEP {\bf 1210}, 066 (2012).

S.~Bhowmick and S.~Mukherji,
Mod.\ Phys.\ Lett.\ A {\bf 27}, 1250009 (2012).

  S.~Hervik, D.~F.~Mota and M.~Thorsrud,
  JHEP {\bf 1111}, 146 (2011).



C.~G.~Boehmer, D.~F.~Mota,
Phys.\ Lett.\  {\bf B663}, 168-171 (2008).

T.~S.~Koivisto, D.~F.~Mota,
JCAP {\bf 0808}, 021 (2008).


D.~H.~Lyth and M.~Karciauskas,
JCAP {\bf 1305}, 011 (2013).

Tuan Q. Do and W. F. Kao,
Phys. Rev. D 84, 123009.\\
Tuan Q. Do, W. F. Kao, and Ing-Chen Lin,
Phys. Rev. D 83, 123002.


A.~Naruko, E.~Komatsu and M.~Yamaguchi,
  JCAP {\bf 1504}, no. 04, 045 (2015).






\bibitem{Kanno:2010ab}
M.~a.~Watanabe, S.~Kanno and J.~Soda,
 Prog.\ Theor.\ Phys. \ {\bf 123} 1041 (2010) [arXiv:1003.0056 [astro-ph.CO]].

\bibitem{Emami:2013bk}
  R.~Emami and H.~Firouzjahi,
  JCAP {\bf 1310}, 041 (2013)
  [arXiv:1301.1219 [hep-th]].

\bibitem{Himmetoglu:2008ab}
  A.~E.~Gumrukcuoglu, B.~Himmetoglu, and M.~Peloso,
  Phys.\ Rev.\ D\  {\bf 81}, 063528 (2010)
  [arXiv:1001.4088 [astro-ph]].


\bibitem{Emami:2016ldl}
  R.~Emami, S.~Mukohyama, R.~Namba and Y.~l.~Zhang,
  JCAP {\bf 1703}, no. 03, 058 (2017)
  doi:10.1088/1475-7516/2017/03/058
  [arXiv:1612.09581 [hep-th]].








\end{thebibliography}
\end{document}